\documentclass{sig-alternate}

\usepackage{alltt                                    % I like these
          , multirow
          , booktabs
          , listings
          , graphicx
          ,float
	,cite
          ,verbatim
         ,mathtools
	,url
	,amsmath
}
\usepackage[table]{xcolor}
\usepackage[numbers]{natbib}     % this is a better citation system
\usepackage{syntax}
\usepackage{algorithmic, algorithm}
\usepackage{enumitem}
\usepackage{threeparttable}
\usepackage{pbox}
\usepackage[font=small,skip=1pt]{caption}

\usepackage{expl3}
\ExplSyntaxOn
\newcommand\latinabbrev[1]{
  \peek_meaning:NTF . {% Same as \@ifnextchar
    #1\@}%
  { \peek_catcode:NTF a {% Check whether next char has same catcode as \'a, i.e., is a letter
      #1., \@ }%
    {#1., \@}}}
\ExplSyntaxOff

\newcommand{\CASE}[1]{\STATE \textbf{case} #1\textbf{:} \begin{ALC@g}}
\newcommand{\ENDCASE}{\end{ALC@g}}

\newcommand{\DEFAULT}{\STATE \textbf{default:} \begin{ALC@g}}
\newcommand{\ENDDEFAULT}{\end{ALC@g}}
\newcommand{\DEFAULTLINE}[1]{\STATE \textbf{default:} }
\let\footnotesize\scriptsize

\newsavebox{\supbox}% Superscript box
\newcommand{\bsup}{\begin{lrbox}{\supbox}$\tt\scriptstyle}% Superscript begin
\newcommand{\esup}{$\end{lrbox}{}^{\usebox{\supbox}}}% Superscript end
\def\eg{\latinabbrev{e.g}}
\def\ie{\latinabbrev{i.e}}

\definecolor{lightpurple}{rgb}{0.8,0.8,1}
\definecolor{codebg}{RGB}{255,255,255}
\definecolor{commentcolor}{RGB}{11,140,11}
\begin{document}
%
% --- Author Metadata here ---
\conferenceinfo{MSR}{'2014, Hyderabad, India}
%\CopyrightYear{2007} % Allows default copyright year (20XX) to be over-ridden - IF NEED BE.
%\crdata{0-12345-67-8/90/01}  % Allows default copyright data (0-89791-88-6/97/05) to be over-ridden - IF NEED BE.
% --- End of Author Metadata ---

\title{An Insight into the Pull Requests of GitHub\vspace{-0.3cm}}

\numberofauthors{3} %  in this sample file, there are a *total*
\author{
% 1st. author
%\alignauthor
\begin{tabular}[t]{@{}c@{}}
Mohammad Masudur Rahman~~~~Chanchal K. Roy\\
       \affaddr{University of Saskatchewan, Canada}\\
       \email{\{mor543, ckr353\}@mail.usask.ca}
% 2nd. author
\end{tabular}
% 3rd. author
}

% There's nothing stopping you putting the seventh, eighth, etc.
% author on the opening page (as the 'third row') but we ask,
% for aesthetic reasons that you place these 'additional authors'
% in the \additional authors block, viz.
%\additionalauthors{Additional authors: John Smith (The Th{\o}rv{\"a}ld Group,
%email: {\texttt{jsmith@affiliation.org}}) and Julius P.~Kumquat
%(The Kumquat Consortium, email: {\texttt{jpkumquat@consortium.net}}).}
% Just remember to make sure that the TOTAL number of authors
% is the number that will appear on the first page PLUS the
% number that will appear in the \additionalauthors section.
%StackOverflow, a popular programming Q \& A site, often contains working code examples that solve particular programming problems, as a part of the answers against the posted questions. Studies show that the community is highly interested of those code examples, and the examples contribute greatly to the promotion and demotion of the answers. 
\maketitle
\begin{abstract}
%ROY
%In GitHub, once a developer completes a milestone and makes a pull request, other members of the project analyze the posted commits, and review the associated code. The posted commits are accepted only if both the merge operation succeeds without conflicts and the concerns identified by other developers are properly addressed. 
Given the increasing number of unsuccessful pull requests in GitHub projects, insights into the success and failure of these requests are essential for the developers. In this paper, we provide a comparative study between successful and unsuccessful pull requests made to 78 GitHub base projects by 20,142 developers from 103,192 forked projects. In the study, we analyze pull request discussion texts, project specific information (\eg\ domain, maturity), and developer specific information (\eg\ experience) in order to report useful insights, and use them to contrast between successful and unsuccessful pull requests. We believe our study will help developers overcome the issues with pull requests in GitHub, and project administrators with informed decision making.
\end{abstract}

%A category including the fourth, optional field follows...

%\terms{Theory, Metrics, Human factors}

%\keywords{Commit comments, pull request, topic model }

%%introduction, why developer use working code examples
%%However, the codes by the developer are not regularly or closely monitored by other open source developers unlike in the controlled commercial development, which leaves potential bugs or inefficiencies in the codes unidentified. 
%; however, a formal analysis about the commits is not often done by other developers especially if the merging is successful. Second, when an issue with the code in base repository is found at a later revision after successful merge operation, it becomes a challenge to map the issue with the target commit that introduces it, especially when a number of commits and revisions are associated with the project. The problem is also partially contributed to by the delayed merging process and the lack of closed and frequent monitoring of commit operations.
\section{Introduction}
GitHub, a popular web-based source code hosting service, provides a convenient way for the software developers to collaborate on open source software development with one another. In order to contribute, a developer either creates her own repository or forks from a \emph{base repository}, and continues her work. GitHub maintains the source code and associated content (\eg\ committed code, commit comments) for both base and forked repositories separately. The idea is to allow the developer to continue her work without reporting every single commit instantly to the \emph{base repository}. The approach helps her to avoid the \emph{frequent merge conflicts} with other developers of the project, and also provides flexibility in the development. Once the developer completes a milestone (\eg\ module) involving several commits to the forked repository, she makes a \emph{pull request} to the owner (\ie\ administrator) of the \emph{base repository}, and attempts to get her commits merged. Then other members of the project analyze the posted commits, review the code, and the streams of discussion among them are captured in terms of \emph{pull request commit comments}. The posted commits are generally accepted if both the merge operation succeeds without conflicts and the identified concerns by other developers are properly addressed. Unfortunately, not all the pull requests succeed and there are growing concerns of how to make successful pull requests in GitHub \cite{george}. In this research, we perform a comparative study between successful (\ie\ merged with base repository) and unsuccessful (\ie\ failed to merge with base repository) pull requests by analyzing different related artifacts such as the pull request discussion texts (\ie\ code review comments), pull request history, and project and developer specific statistics. The study can provide important insights into the success and the failure of a pull request at GitHub repositories.
%We usethe corresponding discussion texts (\ie\ code review comments) and apply topic modeling on them to discover the frequent issues associated with the pull requests. 
%Unfortunately, there has been a little attempt to investigate into the pull requests in the code repository.
%In this paper, we conduct an exploratory study applying a machine learning technique on the pull request discussion texts, and we discover the frequent issues and inefficiencies in the source codes hosted at GitHub.

A number of existing studies focus on the analysis of email messages \cite{email}, bug reports \cite{trendy}, MSR papers \cite{happy}, and commit messages of source code repositories \cite{hot, topiclabel} for various software maintenace activities. Our work is closely related with the study by \citet{topiclabel}, where they extract the hidden topics from the commit comments of a code repository, and then map to different cross-project non-functional requirements in order to analyze the software maintenance activities. 
In this paper, we examine the pull request discussion texts along with project and developer specific information using a machine learning technique and then report the frequent technical issues and inefficiencies in the source code hosted at GitHub. We use MSR dataset \cite{george}, and collect information about 78,517 pull requests made to 78 base projects by 20,142 developers from 103,192 forked projects. We extract 100 underlying topics that the reported issues of 9,421 pull requests (containing pull request discussion) are based on. In order to extract the topics, Latent Dirichlet Allocation (LDA) with Gibbs sampling is used, and we
manually label 64 topics. We identify eight frequently discussed technical topics, and manually analyze the pull request discussion texts for useful insights. From the analysis of project and developer specific information, our study reports that programming language and domain specific factors can influence the success and failure rates of the pull requests. More importantly, it finds out that success rate of pull requests for a project degrades comparatively with a large number developers (\eg\ more than 4,000) or a large number of forked projects (\eg\ more than 3,000). While the extracted frequent technical topics and language or domain specific insights can help the developers with successful pull requests, project and developer specific insights can aid the GitHub project administrators with informed decision making in the management of pull requests, projects and developers involved.

%We also attempt to find the correlation between the \emph{success} and the \emph{failure} of a pull request and other factors such as programming language, domain, age and maturity of project, developer experience, and number of developers in a project. 
%Finally, we provide a repository specific analysis about the possible issues, risks and threats for each code repository using their corresponding comments and the trained LDA model, whcih can help in further development and administration of the repository.

\begin{figure}[!t]
\centering
\includegraphics[width=3.32in ]{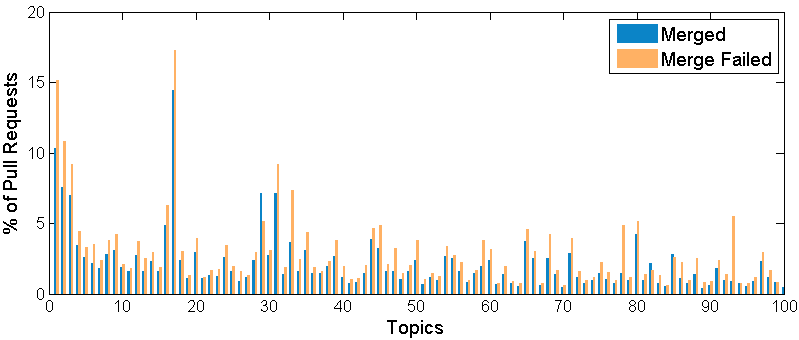}
%\vspace{-.1cm}
\caption{Pull Requests vs. Topics}
\label{fig:topdoccount}
\vspace{-.3cm}
\end{figure}

\section{Dataset}
\label{sec:dataset}
MSR challenge dataset \cite{george} contains 78,955 pull requests (33,910 merged and 45,045 merge failed)  made to 88 \emph{base projects} by 20,142 developers. Among them, 9,601 pull requests (4,091 merged and 5,510 merge failed) made to 78 base projects contain \emph{pull request commit comments}. Pull request commit comments are generally short comments containing code reviews by other developers of the project. Each pull request contains a series of conversations (\eg\ comments), and discusses a few topics concerning the committed code. We use the challenge datatset, and consider the whole sequence of conversations associated with a successful or a failed pull request as a document, and collect 9,601 conversation documents for the experiment. It should be noted that we limit our study to the 78 \emph{base projects} and their forked projects, and also use other 68,916 pull requests (\ie\ not containing discussion) for the experiment. We collect the details of each project (\eg\ domain, programming language, age and maturity) and the corresponding developers (\eg\ number, experience) for the comparative analysis. The hosted projects belong to different application domains such as \emph{reusable frameworks} and \emph{libraries, networking, database management, IDE, statistics} and so on. The projects are written in 13 different programming languages such as \emph{Scala, Python, Java, C\#} and so on.

%The topic model considers a document as a distribution over topics, and a topic as the distribution over a few words \cite{trendy}. It considers the co-occurrence of words within a document as the measure of relatedness, and  provides a probabilistic measure of how the document or the word describes a certain topic.

\section{Tools and Methodology}
\label{sec:tools}
We apply Latent Dirichlet Allocation (LDA) \cite{trendy}, a popular topic modeling technique, on the document collection to find out the underlying topics discussed in each document (\eg\ pull request conversation). In order to apply topic modeling on the corpus, we normalize the content of each document given that they contain natural language texts. We remove stop words from them using a word list\footnote{https://code.google.com/p/stop-words/} provided by Google, and  perform stemming to extract the root form of each word. Then we represent the content of each document as a collection of stemmed tokens, and we use 9,421 such documents. It should be noted that after stemming and removal of stop words, we find the content of 180 documents insignificant (\eg\ empty, contains single word), and they are discarded from the corpus. 

\textbf{Topic Modeling:} We use \emph{JGibbLDA}\footnote{http://jgibblda.sourceforge.net/}, a LDA implementation that uses Gibbs sampling, in order to determine the probabilistic topic model. We use 3,000 iterations of sampling, and collect 100 topics discussed in the document collection, where each topic is described using ten relevant words. We also collect the probabilistic measures of the extent to which a corpus document discusses each of the 100 topics, which we use for comparative analysis in the later phase.

\textbf{Topic Labeling:} The tool returns a list of ten relevant words along with their probabilistic expressiveness for each topic. However, the extracted topics should be more comprehensive for effective analysis which demands topic labeling. In order to label a topic, we analyze the corresponding word list and choose the top four words representing the topic. These words are not necessarily the words with the topmost probabilities. The labeling approach is partially motivated by the approach of \citet{autolabel}, where they use the article titles extracted from \emph{Wikipedia} search for automatic topic labeling. In our research, we use the most representative words from the list and a few programming or technical jargons extracted from various online sources\footnote{http://www.webopedia.com/Programming}, and manually label each topic. We successfully label 64 out of 100 topics, and the labeled topics are hosted elsewhere \cite{expdata} due to space limitation. We show the top four representative words for each of the topics in the table, and the complete word list for each of the topics can also be found online \cite{expdata}.

\begin{figure}[!t]
\centering
\includegraphics[width=3.32in ]{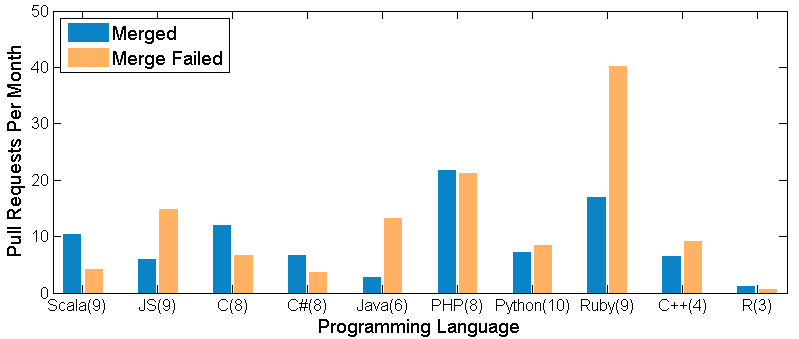}
\caption{Pull Requests vs. Languages\vspace{-.3cm}}
\label{fig:plcount}
\vspace{-.3cm}
\end{figure}
\section{Comparative Analysis}
In this study, we analyze different artifacts and aspects associated with the pull requests in order to extract useful information that can provide insights into the success and failure of the pull requests. 

\textbf{Technical Topics Discussed}: The discussion texts associated with a pull request often contain useful information about the technical concerns identified in the code. In our research, we identify and use them to contrast  between successful and failed pull requests. From \emph{JGibbLDA} tool, we collect the probabilistic alignments of a document (\eg\ pull request discussion) to each of the 100 extracted topics. We sort the measures in descending order and collect the top five dominant topics (\ie\ due to short volume of each discussion) discussed in the document. Then for each topic, we collect the frequency of the pull requests (\ie\ discussion documents) and the information regarding their successes and failures. Fig. \ref{fig:topdoccount} shows the percentage of the 9,421 pull requests that involve each topic in the discussion. We note that eight topics are widely discussed (\ie\ each topic on average in 17.54\% documents), and six of them can be labeled-- \emph{Recursion and Refactoring} (in 18.35\% documents), \emph{Database Query Execution} (in 16.16\% documents), \emph{Arrays and functions} (in 11.12\% documents), \emph{Actor Model} (in 12.22\% documents), \emph{OOP Paradigm} (in 16.29\% documents) and \emph{Space and Indents} (in 10.98\% documents). The two unlabeled dominant topics are discussed in 28.60\% documents on average. The rest 92 topics are less discussed (\ie\ each topic in less than 4\% documents on average). We also note that each topic is \emph{more prevalent} in the discussion of the unsuccessful pull requests than that of the successful pull requests except a dominant topic-- \emph{Actor Model}. Thus, the study reports that a few technical problems (\ie\ topics) are frequently faced by the developers in the pull requests; however,  most of the time, they fail to address them, and therefore, the pull requests do not succeed.

\textbf{Programming Language}: Programming language of a project is an important aspect to take into account when we are interested in comparative analysis of pull requests. We find 13 programming languages used in the 78 GitHub projects, and we find nine of them having 8-10 projects each. However, we also select \emph{R} language containing three projects, and discard \emph{CSS}, \emph{Go} and \emph{TypeScript} from the experiment due to their insignificant number of projects. We consider the average number of successful and failed pull requests for a project from each of the programming languages. We also note that age of the project can be an influencing factor in this case, and therefore, we determine the average number of successful and unsuccessful pull requests made per month. Fig. \ref{fig:plcount} shows the average number of pull requests made per month to any single base project of each programming language by its forked projects. We note that projects using \emph{Scala, C, C\#, R} and \emph{PHP} programming languages made more successful pull requests on average than failed ones, whereas projects of \emph{JavaScript(JS), Java, Python, Ruby} and \emph{C++} did the opposite. We investigate into the forks and the developer pool associated with those projects, and find out that the first group of projects except the \emph{PHP}-based ones have less forks but more developers involved than those of the later group. We also note that on average, \emph{PHP} and \emph{Ruby}-based projects made the highest number of pull requests per month; however, Ruby-based projects are often found with increasing number of failed pull requests per month. Although we speculate, this is due to the maximum number of forks in \emph{Ruby} projects, the finding can encourage the research on the language specific factors on pull requests.

%Thus, the findings validate our intuition that programming language paradigm can be a influential factor to the success or failure of a pull request.

\begin{figure}[!t]
\centering
\includegraphics[width=3.2in ]{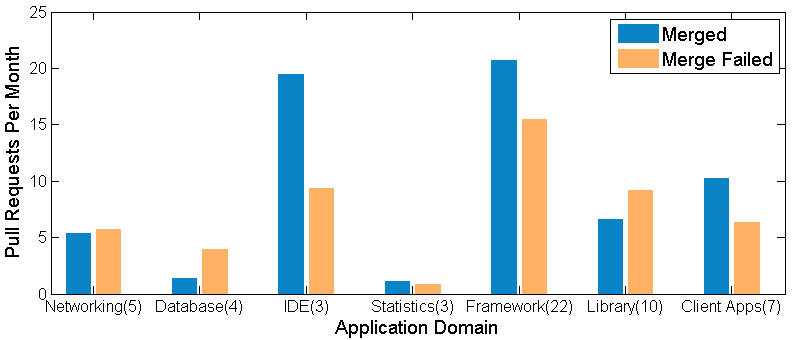}
\caption{Pull Requests vs. Domains}
\label{fig:domain}
\vspace{-.3cm}
\end{figure}

\textbf{Application Domain:} Domain specific concerns can be introduced in the pull request discussions, and they can affect the chance of merging for a pull request. In our research, we consider the domain of the project, and  identify seven major domains-- \emph{Networking, Database, IDE, Statistics, Framework, Library} and \emph{Client Apps}. We manually categorize each of the 78 base projects into different domains consulting their documentations provided online. We also determine the number of pull requests received each month by a \emph{base project} from a particular domain. Fig. \ref{fig:domain} shows the pull request statistics for each domain. We note that projects from \emph{Framework} and \emph{IDE} domains made the maximum number of pull requests each month, and their success rates are relatively higher than that of the projects from other domains. Projects from \emph{Networking, Library} and \emph{Client apps} domains showed comparatively similar success and failure rates in the merging of pull requests.

\begin{figure}[!t]
\centering
\includegraphics[width=3.2in ]{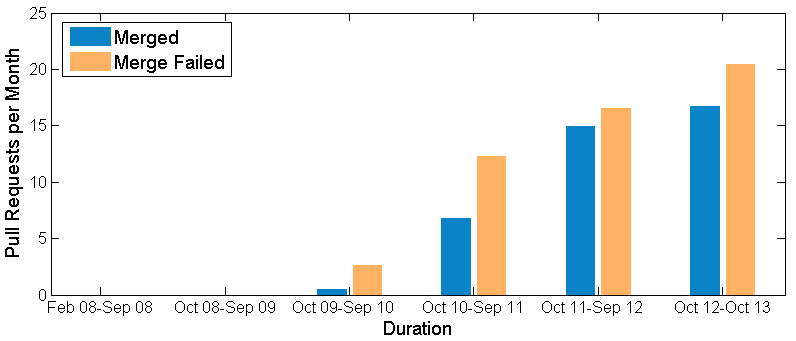}
\caption{Pull Requests vs. Durations}
\label{fig:duration}
\vspace{-.3cm}
\end{figure}
\textbf{Project Age \& Maturity}: Over time, a project either may get lost or get matured through the collaboration of a number of open source developers online. We believe that \emph{age} (\ie\ the time interval between project creation date and latest recorded date in the dataset, October 5, 2013) and \emph{number of forked projects} are two important factors that may influence the success rate of the pull requests of a project. We consider a timeline of five plus years from February, 2008 to October, 2013 with one year interval, and determine the average number of pull requests made to any single base project each month during each interval. Fig. \ref{fig:duration} shows the results of the experiment. We note that up to September, 2009, no pull request are made, and from October, 2009 to onward, the pull requests (\ie\ both successful and failed) increase almost exponentially. It should be noted that throughout the intervals there is an increasing trend on age and number of projects, and developer participation, which actually help the higher rate of pull requests for the projects.

\begin{figure}[!t]
\centering
\includegraphics[width=3.2in ]{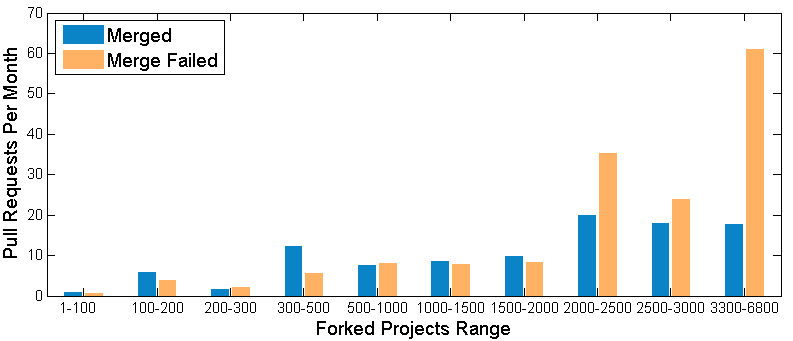}
\caption{Pull Requests vs. Project Maturity}
\label{fig:maturity}
\vspace{-.3cm}
\end{figure}

We consider the number of forked projects as an estimate of the maturity of a base project. We find at most 103,192 forks for 78 base projects, and we choose certain ranges. Then, for each fork count range, we determine the average number of pull requests made to a corresponding base project each month. Fig. \ref{fig:maturity} shows the results of the experiment. We note that with increase in forked projects, the average number of pull requests per month increases; however, it does not show a regular pattern. Moreover, with the increase in forked projects, the failure rate of pull requests increases especially for the projects with more than 2,000 forks.

\begin{figure}[!t]
\centering
\includegraphics[width=3.2in ]{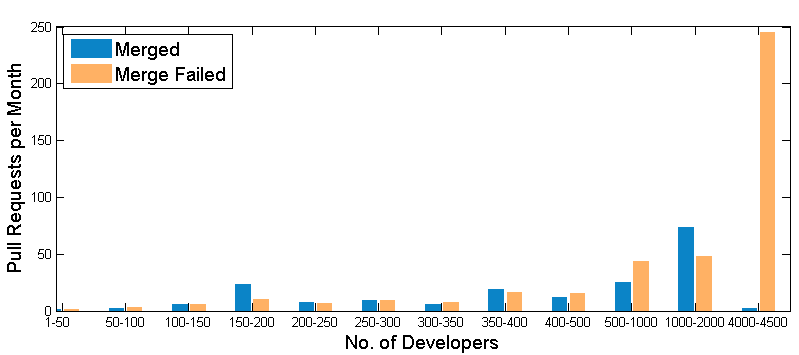}
\caption{Pull Requests vs. No. of Developers}
\label{fig:numdev}
\vspace{-.3cm}
\end{figure}

\begin{figure}[!t]
\centering
\includegraphics[width=3.2in ]{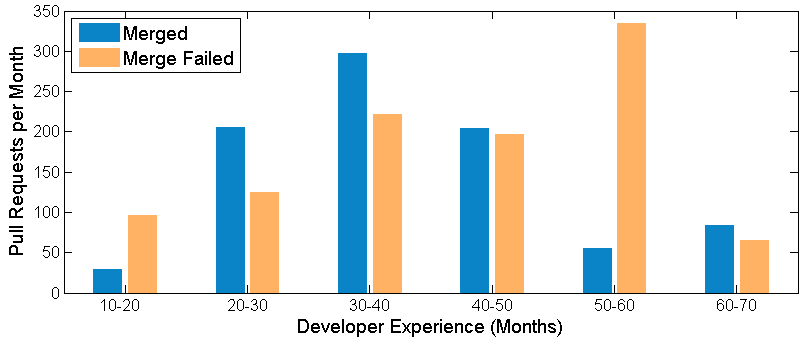}
\caption{Pull Requests vs. Experience}
\label{fig:devexp}
\vspace{-.3cm}
\end{figure}
\textbf{Project Developers \& Experience}: Number of developers involved into a project along with their working experience with the project are also two contributing factors that can influence the success and failure rate of the pull requests. We collect the information of 20,142 developers involved into 78 base projects, whose working experience varies from five months to 68 months. We sort the projects according to their developer range, and Fig. \ref{fig:numdev} shows the average number of pull requests made to a base project each month with a certain range of developers. We note that the average number of pull requests does not increase comparatively with higher participation; however, the projects with a large developer crowd may make an excessive number of unsuccessful pull requests.

We choose a set of ranges for the developer experience, and Fig. \ref{fig:devexp} shows the average number of pull requests made each month to a base project having developers in the forked projects with certain range of experience. We interestingly note that developers with 20 months to 50 months of experience are found the most productive, and they made the maximum number of pull requests each month. However, developers with further experiences are found either less productive or making a number of unsuccessful pull requests each month. We can speculate that the experienced developers might be involved in management activities rather development; however, more insights could be extracted if the role information is provided, which the dataset does not provide.  

\section{Discussion and Conclusion}

In this study, we conduct a comparative study between successful and unsuccessful pull requests made to 78 GitHub base projects by 20,142 developers from 103,192 forked projects. We analyze the pull request discussion texts that contain useful information about the frequent technical issues encountered. We also analyze the pull request history along with project and developer specific information in order to extract useful insights into the success and the failure of pull requests. This section reports our findings in brief as follows:

Eight \emph{topics} (\ie\ technical issues), six of which can be labeled--\emph{Recursion and Refactoring}, \emph{Database Query Execution}, \emph{Arrays and functions}, \emph{Actor Model}, \emph{OOP Paradigm} and \emph{Space and Indents}, are widely discussed in the texts of pull request discussion. Each of those eight technical issues is faced by 17.54\% of the 9,421 pull requests on average, and 7.76\% of the requests succeed to merge. The rest 92 topics are less discussed, and each of them is covered only in 3.90\% of the discussion.

In case of 24 GitHub \emph{base projects} using three \emph{programming languages}--\emph{Ruby, Java} and \emph{JavaScript}, average number of unsuccessful pull requests per month is exceptionally higher than that of successful pull requests. While the study points out \emph{excessive forking} as a possible reason, future research can investigate into the language specific factors affecting the success and failure of the pull requests.

Most of the projects under study belong to seven major \emph{application domains}, and three of them--\emph{Framework, Library} and \emph{Client Apps}, contain 39 projects. Projects from \emph{IDE, Framework} and \emph{Client Apps} domains demonstrate a comparatively higher success rate of pull requests, whereas projects from \emph{Database} and \emph{Statistics} domains show very limited activity in terms of pull requests. The study suggests the \emph{limited number of projects} in the target domains as a possible explanation of low activity; however, it also encourages the future research on the domain specific concerns affecting pull requests.   

The \emph{age} and \emph{maturity} (\ie\ number of forks) of a GitHub project clearly affects the success and failure rates of pull requests. As time goes by and more forks are created from the base project, the number of pull requests increases and so do their success and failure rates. More importantly, our study finds that the failure rate of pull requests increases rapidly (for seven projects) when more than 3000 forks are created, which can be an important piece of information for the administrators of the \emph{base projects}.

The \emph{number of developers} involved into a project and their \emph{experience} can affect the success and failure rates of pull request for a project. Our study finds that the average number of pull requests per month for a project increases almost regularly against increased participation of the developers; however, the rate of unsuccessful pull requests increases exponentially (for one project) with more than 4000 developers involved. The study also suggests that developers with 20 to 50 months experience are found the most productive in terms of submitting and getting pull requests accepted. While we cannot provide a suitable explanation for this due to lack of information in the dataset, the findings can help the project administrator to attract the appropriate audience, and manage the existing developer pool involved into the project.

%Roy
%\section{Conclusion}
%\label{sec:conclusion}
%To summarize, in this research, we conduct a comparative study between successful and unsuccessful pull requests made to 78 GitHub base projects by 20,142 developers from 103,192 forked projects. We analyze the pull request discussion texts that contain useful information about the frequent technical issues encountered. We also analyze other related aspects such as programming language, application domain, age, and maturity of the projects, and the developer crowd contributing to the projects. The work provides important insights into the success and failure of the pull requests which can help the open source developers overcome the technical issues associated with pull requests, and project administrators with informed decision making in the management of projects and developer pool involved. In future, we plan to conduct the study with an extended dataset.
\vspace{-.18cm}

\bibliographystyle{plainnat}
\setlength{\bibsep}{0pt plus 0.3ex}
\scriptsize
\bibliography{sigproc}  % sigproc.bib is the name of the Bibliography in this case

\begin{thebibliography}{8}
\providecommand{\natexlab}[1]{#1}
\providecommand{\url}[1]{\texttt{#1}}
\expandafter\ifx\csname urlstyle\endcsname\relax
  \providecommand{\doi}[1]{doi: #1}\else
  \providecommand{\doi}{doi: \begingroup \urlstyle{rm}\Url}\fi

\bibitem[exp()]{expdata}
{E}xperiment {D}ata.
\newblock URL \url{http://www.usask.ca/~mor543/msr2014}.

\bibitem[Demeyer et~al.(2013)Demeyer, Murgia, Wyckmans, and Lamkanfi]{happy}
S.~Demeyer, A.~Murgia, K.~Wyckmans, and A.~Lamkanfi.
\newblock {H}appy {B}irthday! {A} {T}rend {A}nalysis on {P}ast {MSR} {P}apers.
\newblock In \emph{Proc. MSR}, pages 353--362, 2013.

\bibitem[Dredze et~al.(2008)Dredze, Wallach, Puller, and Pereira]{email}
M.~Dredze, H.M. Wallach, D.~Puller, and F.~Pereira.
\newblock {G}enerating {S}ummary {K}eywords for {E}mails {U}sing {T}opics.
\newblock In \emph{Proc. IUI}, pages 199--206, 2008.

\bibitem[Gousios(2013)]{george}
G.~Gousios.
\newblock {T}he {GHT}orrent {D}ataset and {T}ool {S}uite.
\newblock In \emph{Proc. MSR}, pages 233--236, 2013.

\bibitem[Hindle et~al.(2009)Hindle, Godfrey, and Holt]{hot}
A.~Hindle, M.W. Godfrey, and R.C. Holt.
\newblock {W}hat's {H}ot and {W}hat's {N}ot: {W}indowed {D}eveloper {T}opic
  {A}nalysis.
\newblock In \emph{Proc. ICSM}, pages 339--348, 2009.

\bibitem[Hindle et~al.(2011)Hindle, Ernst, Godfrey, and Mylopoulos]{topiclabel}
A.~Hindle, N.A. Ernst, M.W. Godfrey, and J.~Mylopoulos.
\newblock {A}utomated {T}opic {N}aming to {S}upport {C}ross-{P}roject
  {A}nalysis of {S}oftware {M}aintenance {A}ctivities.
\newblock In \emph{Proc. MSR}, pages 163--172, 2011.

\bibitem[Lau et~al.(2011)Lau, Grieser, Newman, and Baldwin]{autolabel}
J.H. Lau, K.~Grieser, D.~Newman, and T.~Baldwin.
\newblock {A}utomatic {L}abelling of {T}opic {M}odels.
\newblock In \emph{Proc.HLT}, pages 1536--1545, 2011.

\bibitem[Martie et~al.(2012)Martie, Palepu, Sajnani, and Lopes]{trendy}
L.~Martie, V.K. Palepu, H.~Sajnani, and C.~Lopes.
\newblock {T}rendy {B}ugs: {T}opic {T}rends in the {A}ndroid {B}ug {R}eports.
\newblock In \emph{Proc. MSR}, pages 120--123, 2012.

\end{thebibliography}


\begin{thebibliography}{15}
\providecommand{\natexlab}[1]{#1}
\providecommand{\url}[1]{\texttt{#1}}
\expandafter\ifx\csname urlstyle\endcsname\relax
  \providecommand{\doi}[1]{doi: #1}\else
  \providecommand{\doi}{doi: \begingroup \urlstyle{rm}\Url}\fi

\bibitem[exp()]{expdata}
Stacksuggest experimental data.
\newblock URL \url{homepage.usask.ca/~masud.rahman/ss/expdata}.

\bibitem[Buse and Weimer(2010)]{readability}
R.P.L. Buse and W.R. Weimer.
\newblock Learning a metric for code readability.
\newblock \emph{Softw. Eng., IEEE Trans.}, 36\penalty0 (4):\penalty0 546--558,
  2010.

\bibitem[Chawla and Chhabra(2012)]{ijert}
Mandeep~K. Chawla and Indu Chhabra.
\newblock Implementing source code metrics for software quality analysis.
\newblock \emph{IJERT}, 1\penalty0 (5), July 2012.

\bibitem[Knight and Myers(1991)]{readuse}
John~C. Knight and E.~Ann Myers.
\newblock Phased inspections and their implementation.
\newblock \emph{SIGSOFT Softw. Eng. Notes}, 16\penalty0 (3):\penalty0 29--35,
  July 1991.

\bibitem[Le~Goues and Weimer(2012)]{specmining}
C.~Le~Goues and W.~Weimer.
\newblock Measuring code quality to improve specification mining.
\newblock \emph{Softw. Eng., IEEE Trans.}, 38\penalty0 (1):\penalty0 175--190,
  2012.

\bibitem[Lochmann and Heinemann(2011)]{lochmann}
Klaus Lochmann and Lars Heinemann.
\newblock Integrating quality models and static analysis for comprehensive
  quality assessment.
\newblock In \emph{Proc. of WETSoM}, pages 5--11, 2011.

\bibitem[M\"{a}ntyl\"{a} and Lassenius(2006)]{subjective}
Mika~V. M\"{a}ntyl\"{a} and Casper Lassenius.
\newblock Subjective evaluation of software evolvability using code smells: An
  empirical study.
\newblock \emph{Empirical Softw. Engg.}, 11\penalty0 (3):\penalty0 395--431,
  September 2006.

\bibitem[Meirelles et~al.(2010)Meirelles, Santos, Miranda, Kon, Terceiro, and
  Chavez]{foss}
P.~Meirelles, C.~Santos, J.~Miranda, F.~Kon, A.~Terceiro, and C.~Chavez.
\newblock A study of the relationships between source code metrics and
  attractiveness in free software projects.
\newblock In \emph{Proc. SBES}, pages 11--20, 2010.

\bibitem[Nasehi et~al.(2012)Nasehi, Sillito, Maurer, and Burns]{nasehi}
S.~M. Nasehi, J.~Sillito, F.~Maurer, and C.~Burns.
\newblock What makes a good code example -a study of programming q and a in
  stackoverflow.
\newblock In \emph{Proc. ICSM}, pages 25 --35, 2012.

\bibitem[Posnett et~al.(2011)Posnett, Hindle, and Devanbu]{simpler}
Daryl Posnett, Abram Hindle, and Premkumar Devanbu.
\newblock A simpler model of software readability.
\newblock In \emph{Proc. MSR}, pages 73--82, 2011.
\newblock ISBN 978-1-4503-0574-7.

\bibitem[Rawat et~al.(2012)Rawat, Mittal, and Dubey]{survey}
Mrinal~Singh. Rawat, Arpita Mittal, and Sanjay~Kumar Dubey.
\newblock Survey on impact of software metrics on software quality.
\newblock \emph{IJACSA}, 3\penalty0 (1), 2012.

\bibitem[Stamelos et~al.(2002)Stamelos, Angelis, Oikonomou, and Bleris]{oss}
Ioannis Stamelos, Lefteris Angelis, Apostolos Oikonomou, and Georgios~L.
  Bleris.
\newblock Code quality analysis in open source software development.
\newblock \emph{Info Systems J}, 12:\penalty0 43--60, 2002.

\bibitem[Stroggylos and Spinellis(2007)]{refactoring}
K.~Stroggylos and D.~Spinellis.
\newblock Refactoring--does it improve software quality?
\newblock In \emph{In Proc. WoSQ}, pages 10--, 2007.

\bibitem[Taibi(2013)]{reusability}
Fathi Taibi.
\newblock Reusability of open-source program code: a conceptual model and
  empirical investigation.
\newblock \emph{SIGSOFT Softw. Eng. Notes}, 38\penalty0 (4), July 2013.

\bibitem[Treude et~al.(2011)Treude, Barzilay, and Storey]{nier}
Christoph Treude, Ohad Barzilay, and Margaret-Anne Storey.
\newblock How do programmers ask and answer questions on the web? (nier track).
\newblock In \emph{ICSE}, pages 804--807, 2011.

\end{thebibliography}


\begin{thebibliography}{8}
\providecommand{\natexlab}[1]{#1}
\providecommand{\url}[1]{\texttt{#1}}
\expandafter\ifx\csname urlstyle\endcsname\relax
  \providecommand{\doi}[1]{doi: #1}\else
  \providecommand{\doi}{doi: \begingroup \urlstyle{rm}\Url}\fi

\bibitem[Buse and Weimer(2010)]{readability}
R.P.L. Buse and W.R. Weimer.
\newblock {L}earning a {M}etric for {C}ode {R}eadability.
\newblock \emph{TSE}, 36\penalty0 (4):\penalty0 546--558, 2010.

\bibitem[Le~Goues and Weimer(2012)]{specmining}
C.~Le~Goues and W.~Weimer.
\newblock {M}easuring {C}ode {Q}uality to {I}mprove {S}pecification {M}ining.
\newblock \emph{TSE}, 38\penalty0 (1):\penalty0 175--190, 2012.

\bibitem[Lochmann and Heinemann(2011)]{lochmann}
K.~Lochmann and L.~Heinemann.
\newblock {I}ntegrating {Q}uality {M}odels and {S}tatic {A}nalysis for
  {C}omprehensive {Q}uality {A}ssessment.
\newblock In \emph{Proc. of WETSoM}, pages 5--11, 2011.

\bibitem[M\"{a}ntyl\"{a} and Lassenius(2006)]{subjective}
M.V. M\"{a}ntyl\"{a} and C.~Lassenius.
\newblock {S}ubjective {E}valuation of {S}oftware {E}volvability using {C}ode
  {S}mells: {A}n {E}mpirical {S}tudy.
\newblock \emph{ESE}, 11\penalty0 (3):\penalty0 395--431, 2006.

\bibitem[Nasehi et~al.(2012)Nasehi, Sillito, Maurer, and Burns]{nasehi}
S.~M. Nasehi, J.~Sillito, F.~Maurer, and C.~Burns.
\newblock What {M}akes a {G}ood {C}ode {E}xample -{A} {S}tudy of {P}rogramming
  {Q} and {A} in {S}tackoverflow.
\newblock In \emph{Proc. ICSM}, pages 25 --35, 2012.

\bibitem[Rawat et~al.(2012)Rawat, Mittal, and Dubey]{survey}
M.S. Rawat, A.~Mittal, and S.K. Dubey.
\newblock {S}urvey on {I}mpact of {S}oftware {M}etrics on {S}oftware {Q}uality.
\newblock \emph{IJACSA}, 3\penalty0 (1), 2012.

\bibitem[Taibi(2013)]{reusability}
F.~Taibi.
\newblock Reusability of {O}pen-{S}ource {P}rogram {C}ode: {A} {C}onceptual
  {M}odel and {E}mpirical {I}nvestigation.
\newblock \emph{SE Notes}, 38\penalty0 (4), 2013.

\bibitem[Treude et~al.(2011)Treude, Barzilay, and Storey]{nier}
C.~Treude, O.~Barzilay, and M.A. Storey.
\newblock How {D}o {P}rogrammers {A}sk and {A}nswer {Q}uestions on the {W}eb?
  ({NIER} {T}rack).
\newblock In \emph{ICSE}, pages 804--807, 2011.

\end{thebibliography}
% ACM needs 'a single self-contained file'!
%\balancecolumns
\end{document}